# A black phosphorus photo-detector for multispectral, high-resolution imaging

Michael Engel[1], Mathias Steiner[1,2,*], and Phaedon Avouris[1,**]

[*]msteine@us.ibm.com , [**]avouris@us.ibm.com

[1] *IBM Thomas J. Watson Research Center, Yorktown Heights, NY 10598, USA*

[2] *IBM Research-Brazil, Rio de Janeiro, RJ 22290-240, Brazil*

**Black phosphorus is a layered semiconductor that is intensely researched in view of applications in optoelectronics. In this Letter, we investigate a multi-layer black phosphorus photo-detector that is capable of acquiring high-contrast (V>0.9) images both in the visible ($\lambda_{VIS}$=532nm) as well as in the infrared ($\lambda_{IR}$=1550nm) spectral regime. In a first step, by using photocurrent microscopy, we map the active area of the device and we characterize responsivity and gain. In a second step, by deploying the black phosphorus device as a point-like detector in a confocal microsope setup, we acquire diffraction-limited optical images with sub-micron resolution. The results demonstrate the usefulness of black phosphorus as an optoelectronic material for hyperspectral imaging applications.**

Among the various 2-dimensional inorganic materials with their unique electronic and optical properties[1], the layered semiconductor black phosphorus is currently intensely researched for its potential applications in electronics[2-5] and optoelectronics[6,7]. The band gap energy of the bulk is 0.3eV[8,9], however, it increases with decreasing layer number; to 1.5eV for a single layer[2]. Therefore, a multi-layer black phosphorus device should, in principle, allow photo-detection over a broad spectral. In this letter, we investigate the photo-response of a multi-layer black phosphorus photo-detector and we demonstrate that it can be used for high-contrast, diffraction-limited optical imaging in the visible as well as the infrared spectral domain.

Black phosphorus is an elemental semiconductor; a stable allotrope of phosphorus[10]. At ambient conditions, orthorhombic black phosphorus forms puckered layers which are held together by van-der-Waals forces[11]. The layered structure allows for mechanical exfoliation to prepare thin layers on a substrate[12]. We have exfoliated multi-layer structures from crystals of black phosphorus on $Si/SiO_2$ substrates and fabricated optoelectronic devices that are suited for photo-detection (for details, see Methods Summary).

In Fig.1, we characterize such a black phosphorus device by laser-excited photocurrent microscopy. The black phosphorus multi-layer has a thickness of 120nm (see atomic force microscope measurement and Raman spectrum of the device in supplemental Figs.S1) and is contacted by metallic leads so that the photocurrent can be extracted and collected by the measurement system. In the actual configuration, the left contact in Fig.1a is connected to the measuring unit while the right contact is grounded. A focused laser beam is raster-scanned across the device by a piezo-driven mirror to acquire photocurrent maps. We overlay the measured short-circuit photocurrent maps excited at $\lambda_{VIS}$=532nm (see Fig.1a) and $\lambda_{IR}$=1550nm (see Fig.1b) onto the reflected laser light images of the device. To analyze the spatial distribution of the photocurrent signal, we perform a cross section of the photocurrent map (along the green arrow, see Fig.1a) which is shown in Fig.1c. The two photocurrent maxima occur in close proximity of the opposing contacts while the photocurrent changes polarity at the inflection point in the middle of the device. Independently of the wavelength, we find that the measured photocurrent involves transport by holes; in agreement with the electrical transfer characteristics of the device that exhibits p-type conduction behavior (see supplemental Fig.S1d). A detailed study of the mechanism of photocurrent generation in black phosphorus, which will be published elsewhere[13], shows that at low bias photodetection arises from photothermal effects, while at high bias bolometric effects dominate. We next quantify the responsivity of our device at $\lambda_{VIS}$ and $\lambda_{IR}$. To that end, we map the photocurrent at different optical excitation power densities and extract the maximum photocurrent amplitude from each measurement, see Fig.1d. We find that the device responsivity is constant within the experimental uncertainty, about 20mA/W at $\lambda_{VIS}$=532nm and 5mA/W at $\lambda_{IR}$=1550nm. Based on the Lambert-Beer law[14,15], by scaling the penetration depth of light to the black phosphorus layer thickness and by assuming comparable light absorption at $\lambda_{IR}$ and $\lambda_{VIS}$, we estimate a responsivity reduction at $\lambda_{IR}$ which is a factor of 3 lower than at $\lambda_{VIS}$. This is in agreement with the experimental data shown in Fig. 2d.

In order to study the gain behavior of the black phosphorus photo-detector, we perform scanning photoconductivity measurements by applying a bias voltage $V_{bias}$ across the device. To distinguish between direct and modulated photocurrent, we chop the laser beam at a frequency $f$=2kHz and measure the AC photocurrent amplitude by a standard lock-in technique. As a result, in Fig.1e, we plot the gain in photoresponse with respect to the unbiased device as a function of bias voltage $V_{bias}$. For both test wavelengths, the gain increases linearly as function $V_{bias}$. The maximum gain obtained at $V_{bias}$=-200mV is 590% at $\lambda_{VIS}$=532nm and 730% and $\lambda_{IR}$=1550nm, respectively. In the following, we will use the measurement conditions $V_{bias}$=0mV with laser light modulation at $f$=2kHz for acquiring images of microscopic structures at $\lambda_{VIS}$=532nm and $\lambda_{IR}$=1550nm.

In Fig.2b, we show scanning electron microscope images of metallic test structures patterned on a glass cover slide which we have used for multispectral optical imaging. In this experiment, the black phosphorus device is deployed as a point-like detector in the image plane of a confocal microscope[16] that is equipped with an immersion objective lens having a numerical aperture (NA) of 1.25 (see schematic in supplemental Fig.S2). The test structures are raster scanned with respect to a tightly focused laser beam at $\lambda_{VIS}$=532nm (see Fig.2c) and $\lambda_{IR}$=1550nm (see Fig.2d), respectively, and the reflected laser light is collected by the black phosphorus photo-detector device (see Fig.1) and converted into electrical current for image formation. The microscope images of the test patterns acquired by the black phosphorus photo-detector are shown in Fig.2c,d. They are diffraction-limited, which we have confirmed by comparing the measured edge cross sections with model point spread functions[16], see Fig. 2e, and theoretical estimates for the resolution limits. At $\lambda_{VIS}$, we obtain an experimental resolution of (270+/-5nm) which agrees with the width of the model point spread function used and an estimate of the theoretical resolution limit[17] of 0.6098· $\lambda_{VIS}$/NA=0.6098·532nm/1.25≈260nm. Likewise, for excitation at $\lambda_{IR}$, we obtain an experimental resolution of (720±15nm) which is in agreement with the width of the model point spread function and the theoretical estimate of 0.6098· $\lambda_{IR}$/NA=0.6098·1550nm/1.25≈760nm. The results demonstrate that it is possible to use a black phosphorus photo-detector for the acquisition of diffraction-limited microscope images in the visible as well as the infrared spectral domain.

For quantifying the effect of imaging wavelength on image resolution, we plot in Fig. 2f the visibility V, or Michelson contrast[18], which is calculated for each image by dividing the difference between measured photocurrent maximum, $I_{max}$, and minimum, $I_{min}$, by their respective sum; $V=(I_{max}-I_{min})/(I_{max}+I_{min})$. As a result, at $\lambda_{VIS}$=532nm, all the image features are fully resolved with high visibility, V=0.93, for all structure sizes shown in Fig. 2c. However, at $\lambda_{IR}$=1550nm, both feature size and feature distance become comparable to or even smaller than the optical resolution limit. We, hence, observe a gradual visibility reduction for smaller feature sizes, from V=0.95 to V=0.79, as expected.

In the following, we investigate the imaging capabilities of the black phosphorus photo-detector with complex text patterns. Fig. 3a shows scanning electron micrographs of text patterns with feature sizes and pitches of 2000nm, 1000nm, and 500nm, respectively. In Fig.3b, we present the optical images measured at $\lambda_{IR}$=1550nm for each test pattern with the black phosphorus device. For the patterns with the feature sizes of 2000nm and 1000nm, the optical images are well-resolved. However, feature sizes of 500nm are below the resolution limit and cannot be fully resolved at $\lambda_{IR}$=1550nm. In order to compare the experimental results quantitatively with theoretical expectations, we first generate binary images (see supplemental Fig.S3) from the scanning electron micrographs shown in Fig.3a. We then perform a two-dimensional convolution of the binary images with a two-dimensional Gaussian point spread function having the width σ. We choose σ to be $\sigma_{IR}$=720nm for $\lambda_{IR}$=1550nm and $\sigma_{VIS}$=260nm for $\lambda_{VIS}$=532nm, respectively, which is consistent with the data shown in Fig.2d. The simulation results are shown in the bottom row of Fig.2b. For each feature size, we find excellent visual agreement between measured and simulated images. The highest microscopic resolution in our experiments with the black phosphorus photo-detector is achieved in the image shown in Fig. 2c. The test patterns with feature sizes and pitches of 500nm are imaged at $\lambda_{VIS}$=532nm and all the features are resolved, in agreement with the theoretical expectation which is represented by the simulated image shown below.

Finally, we would like to point out that the black phosphorus detector has provided a stable performance for all optical excitation powers used in this study. The image acquisition times were chosen identical to those needed for reference measurements with conventional photo-detectors. For example, the integration time per pixel in the image in Fig.3b is 10ms which results in a total image acquisition time of 12min which is a typical value for practical confocal microscopy. During data acquisition, we have not observed noticeable fluctuations of the detector signal and no data post-processing such as averaging or smoothing has been applied to the as-measured images. Taking advantage of detector gain at higher bias, see Fig. 2e, could further improve imaging conditions, depending on the specifics of the application.

In summary, we have fabricated and characterized a photo-detector made of multi-layer black phosphorus and we have used the device to record diffraction-limited images of microscopic patterns in the visible and near-infrared spectral regime. The results demonstrate black phosphorus as a potent material for novel optoelectronics technology, specifically for multi-spectral photo-detection and imaging.

**Methods Summary**

We prepare thin layers of black phosphorus by mechanical exfoliation on highly resistive Si with a 285nm SiO$_2$ layer inside a glove box. Exfoliated layers are then covered with PMMA. We fabricate devices on suitable layers by e-beam lithography which is followed by metal evaporation (1nm Ti, 20nm Pd, 40nm Au) and lift-off in hot acetone. After lift-off, the devices are again covered with PMMA and characterized at room temperature.

**Acknowledgment**

We thank Dr. D. B. Farmer for assistance with sample preparation and valuable discussions, J. J. Bucchignano and B. A. Ek(all IBM Research) for expert technical assistance and Dr. Claudius Feger (IBM Research) for support.

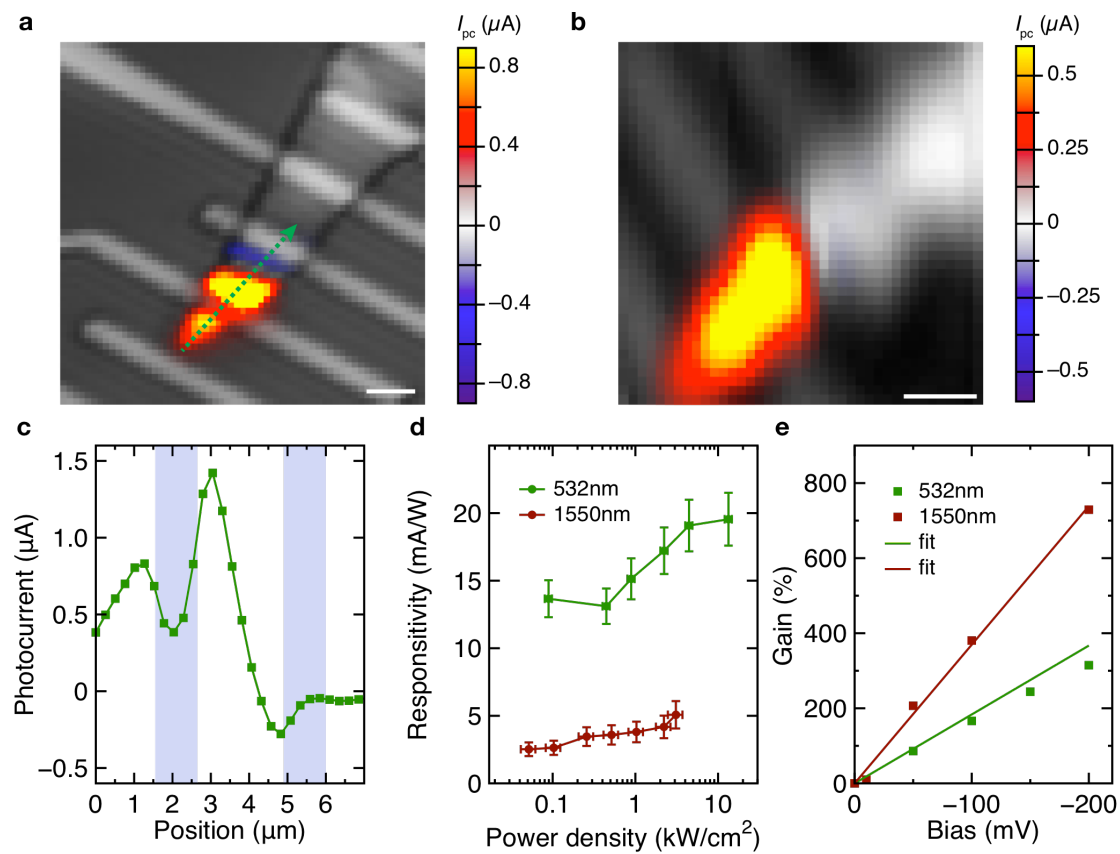

**Figure 1 | Optoelectronic characterization of a black phosphorus photo-detector.** Scanning photocurrent micrographs of the short-circuit photocurrent excited **a** at $\lambda_{VIS}$=532nm with an optical power density of 13.4kW/cm$^2$ and **b** at $\lambda_{IR}$=1550nm with an optical power density of 3.1kW/cm$^2$. The photocurrent maps and the reflected laser light scattering images are overlaid for spatially correlating the photocurrent and the metallic contacts of the device. Scale bar is 2μm. **c** Photocurrent cross section along the green arrow in **a**. Shaded areas indicate the position of the metallic contacts as extracted from the elastic scattering image of the device. **d** Responsivity as function of optical power density for $\lambda_{VIS}$ and $\lambda_{IR}$. For each wavelength, the responsivity values are extracted from the maximum amplitude of each photocurrent map at a given optical power density. **e** Relative gain of the AC photocurrent component at

*f*=2kHz with respect to zero voltage at wavelengths λ$_{VIS}$ and λ$_{IR}$. Linear fits (lines) reveal slopes of 1.84%/mV (λ$_{VIS}$) and 3.69%/mV (λ$_{IR}$).

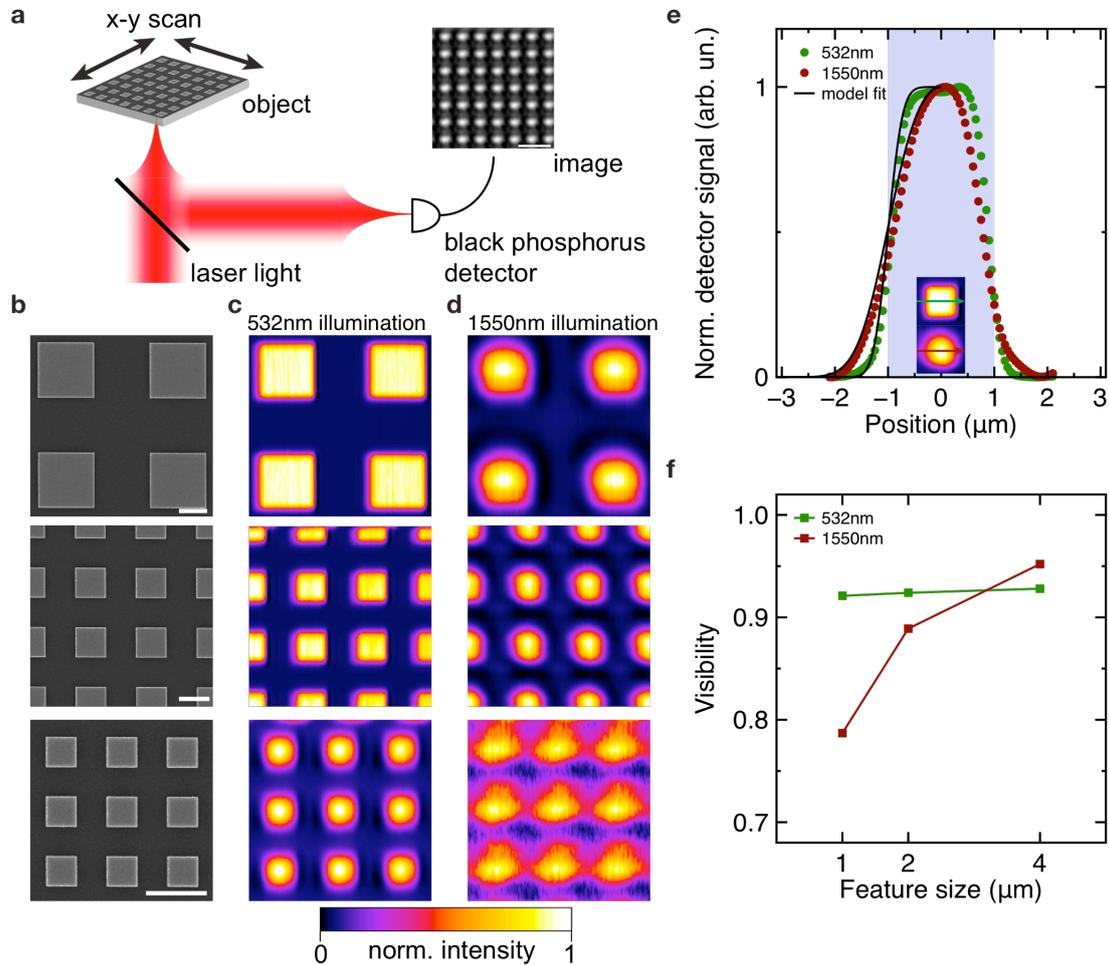

**Figure 2 | Image resolution and visibility of image features recorded with a black phosphorus photo-detector. a** Schematic of the imaging process. The image is an actual measurement of a 500nm square array. Scale bar is 4μm. **b** Scanning electron micrographs of metallic test structures fabricated on glass cover slides having features sizes of 4μm, 2μm, and 1μm. Scale bars are 2μm. Images of test structures excited at **c** λ$_{VIS}$=532nm and **d** λ$_{IR}$=1550nm. Images are the same size as in **b**. **e** Cross sections along the arrows taken from images of a square feature having a side length of 2μm (inset). Images are acquired with a step size of 50nm at wavelengths of λ$_{VIS}$ and λ$_{IR}$. The shaded area indicates the geometry extracted from a scanning electron

micrograph of the feature. The model fit based on a convolution of a step function and a Gaussian function. The optical resolutions extracted from the model functions are (270+/-5nm) and (720±15nm) for $\lambda_{VIS}$ and $\lambda_{IR}$, respectively. **f** Visibility of feature sizes 4μm, 2μm, and 1μm at wavelengths $\lambda_{VIS}$ and $\lambda_{IR}$, extracted from the measured images in **c** and **d**.

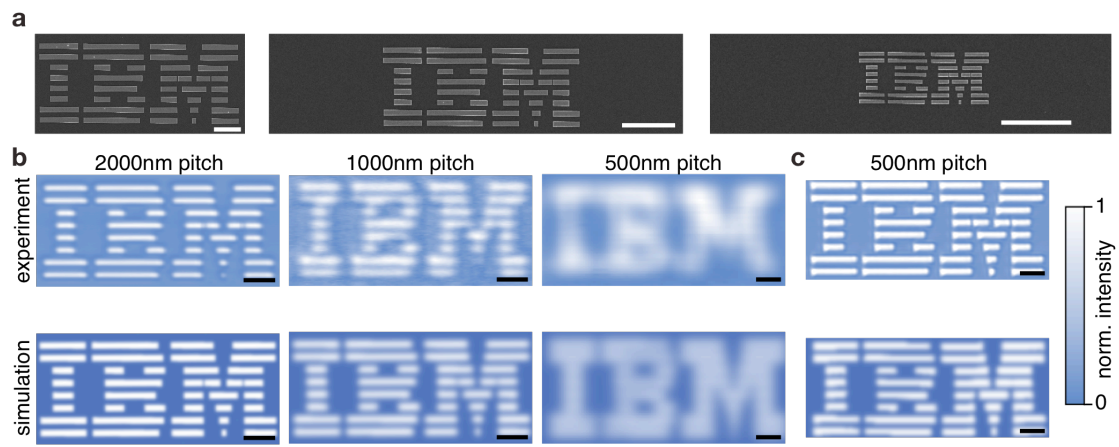

**Figure 3 | Comparison of experimental and simulated microscope images recorded with a black phosphorus photo-detector.** **a** Scanning electron micrographs of test patterns on glass cover slides with feature sizes and pitch of 2000nm, 1000nm, and 500nm. Scale bars are 10μm. **b** Measured images excited at $\lambda_{IR}$=1550nm (top row) and simulated images (bottom row). The simulations are based the scanning electron micrographs shown in **a.** and the optical parameters determined from Fig.2d. Scale bars are 10μm, 5μm, and 2μm. **c** Experimental image excited at $\lambda_{IR}$=532nm (top) exhibiting sub-micron features, along with the simulated image (bottom).

# Supplementary Information:

# A black phosphorus photo-detector for multispectral, high-resolution imaging

Michael Engel[1], Mathias Steiner[1,2,*], and Phaedon Avouris[1,#]

[*]msteine@us.ibm.com , [#]avouris@us.ibm.com

[1] IBM Thomas J. Watson Research Center, Yorktown Heights, NY 10598, USA

[2] IBM Research-Brazil, Rio de Janeiro, RJ 22290-240, Brazil

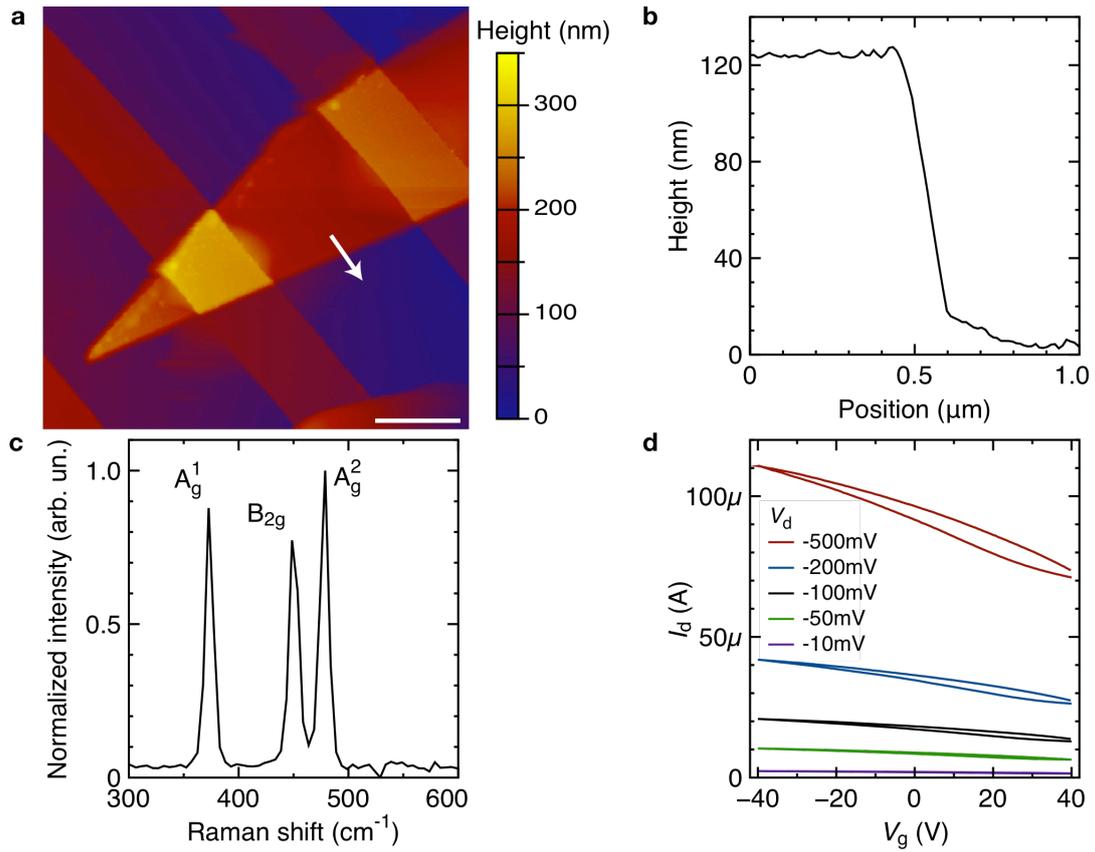

**Figure S1 | Characterization of the black phosphorus photo-detector. a** Atomic force microscope image of the black phosphorus photo-detector discussed in the main manuscript. The scale bar is 2μm. **b** Cross section performed along the direction indicated in by the arrow in **a**. The cross section reveals a layer thickness of about 120nm. **c** Raman spectrum measured on the black phosphorus layer shown in **a**. The Raman spectrum shows the typical low-frequency vibrations of black phosphorus[1]. **d** Electrical transfer characteristics $I_d$-$V_g$ of the device shown in **a** measured for different bias voltages $V_d$.

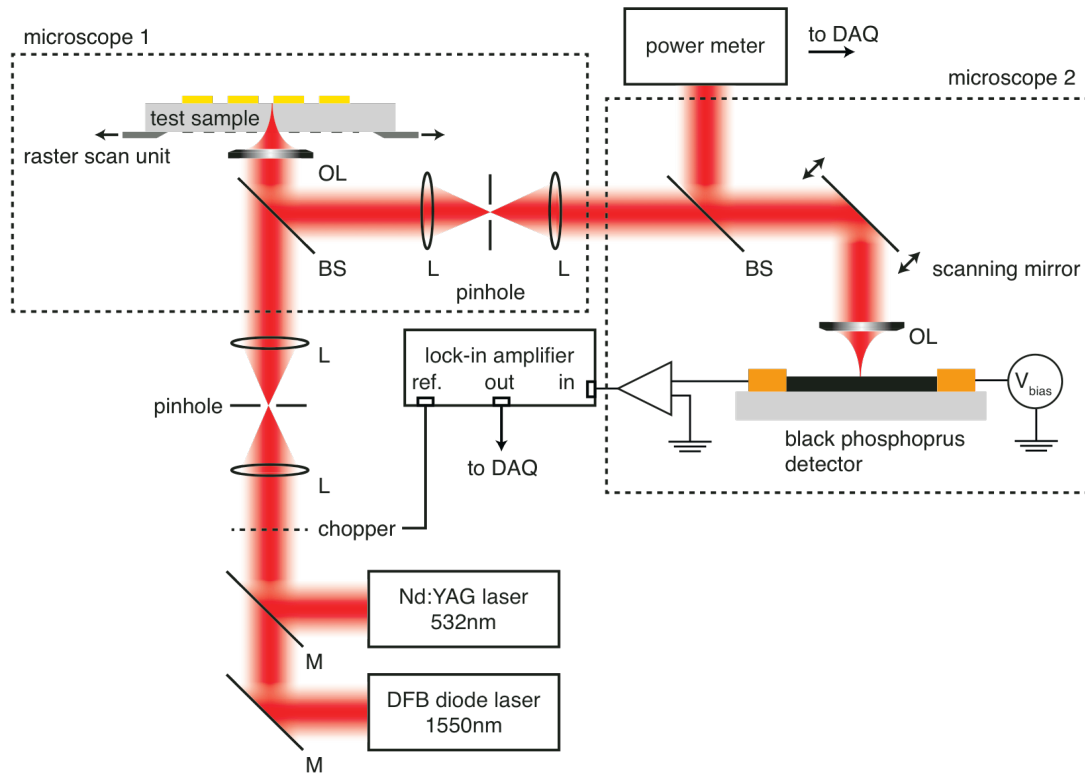

**Figure S2 | Schematic of the experimental setup.** The optical setup combines two confocal microscopes 1, 2 for measuring the microscope images discussed in the main manuscript. For optoelectronic characterization of the black phosphorus photo-detector, microscope 1 is bypassed and laser light is directly coupled into microscope 2. The measured laser-excited photocurrent is amplified and converted into a voltage by a current pre-amplifier. Also, a lock-in amplifier is used to detect the in-phase signal referenced to the mechanically chopped laser beam. For image acquisition with the black phosphorus photo-detector, the test sample is raster scanned with respect to a focused laser beam in microscope 1. The reflected laser light is then directed via a beam splitter into microscope 2 and focused onto the black phosphorus photo-detector.

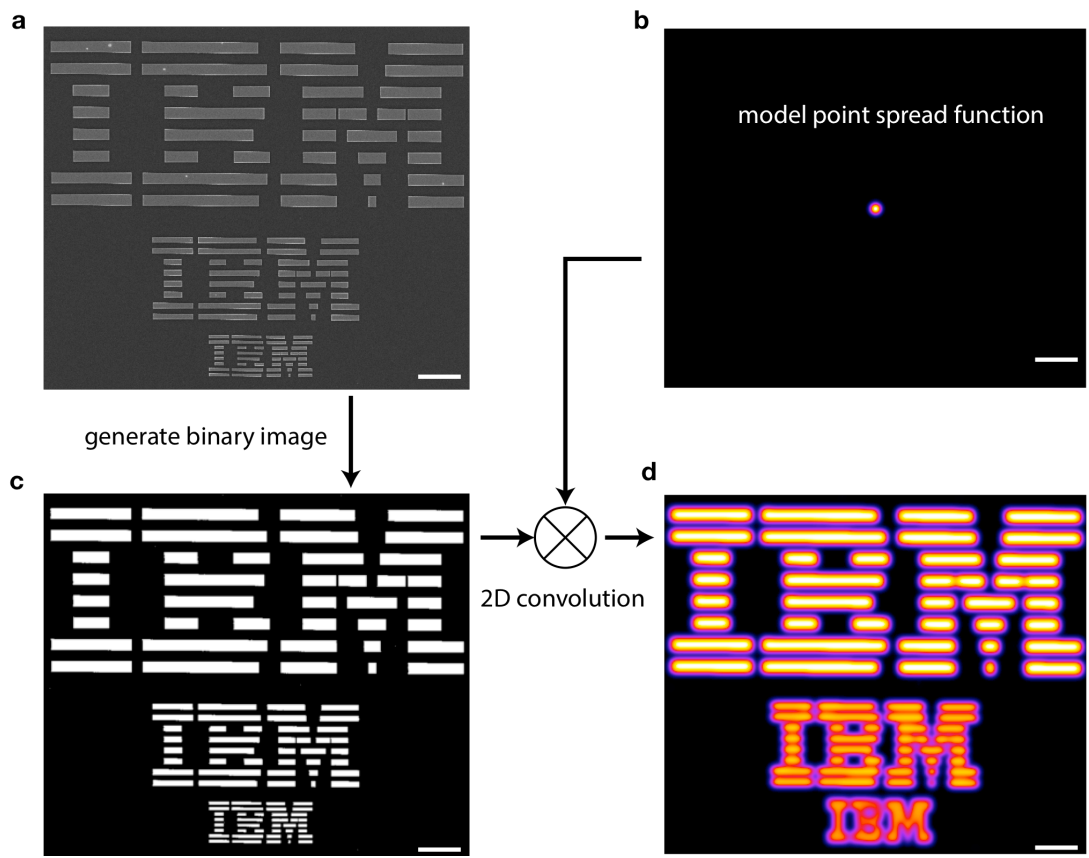

**Figure S3 | Visualization of the image simulation procedure.** In a first step, the scanning electron micrograph **a** is converted into a binary image **b**. In a second step, a model point spread function is generated based on experimentally derived parameters. In a third step, the binary image and the point spread function are convoluted, resulting in a simulated optical image **d**. Scale bars are 8μm. Image transformations and convolutions are performed using ImageJ[2] with the plugin Convolve 3D.